\documentclass[fleqn]{article}
\usepackage{espcrc2}
\usepackage{graphicx}
\usepackage[figuresright]{rotating}
\usepackage{psfig}
\newcommand{\pbar}{\bar{p}}
\newcommand{\ep}{e^+e^- \rightarrow p \overline p}
\newcommand{\pe}{p \overline p \rightarrow e^+e^-}

\newcommand{\bbt}{\bibitem}

\newcommand{\EE}{e^+e^-}
\newcommand{\MM}{\mu^+\mu^-}

\newcommand{\pp}{\pi^+\pi^-}
\newcommand{\kk}{K^+K^-}
\newcommand{\ppb}{p\overline{p}}

\newcommand{\bfg}{\begin{figure}}
\newcommand{\efg}{\end{figure}}
\newcommand{\bitm}{\begin{itemize}}
\newcommand{\eitm}{\end{itemize}}
\newcommand{\bnum}{\begin{enumerate}}
\newcommand{\enum}{\end{enumerate}}
\newcommand{\btbl}{\begin{table}}
\newcommand{\etbl}{\end{table}}
\newcommand{\btbu}{\begin{tabular}}
\newcommand{\etbu}{\end{tabular}}
\begin{document}

\title{ \begin{center} 
Measurement of the cross section for $\ep$ at center-of-mass energies
from 2.0 to 3.07 GeV
\end{center}}
\author{
\begin{center}
M.~Ablikim$^{1}$,              J.~Z.~Bai$^{1}$,               Y.~Ban$^{11}$,
J.~G.~Bian$^{1}$,              X.~Cai$^{1}$,                  H.~F.~Chen$^{16}$,
H.~S.~Chen$^{1}$,              H.~X.~Chen$^{1}$,              J.~C.~Chen$^{1}$,
Jin~Chen$^{1}$
% ...
\\
\vspace{0.2cm}
(BES Collaboration)\\
\vspace{0.2cm}
{\it
$^{1}$ Institute of High Energy Physics, Beijing 100049, People's Republic of China\\
% ...
$^{11}$ Peking University, Beijing 100871, People's Republic of China\\
% ...
$^{16}$ University of Science and Technology of China, Hefei 230026, People's Republic of China\\
% ...
}
\end{center}
}

\begin{abstract}

Cross sections for $\ep$ have been measured at 10
center-of-mass energies from 2.0 to 3.07 GeV by the BESII experiment 
at the BEPC, and proton electromagnetic form factors in the time-like region
have been determined. 

\end{abstract}

\maketitle

\section{Introduction}

Positron-electron annihilation produces hadronic final
states with an amplitude proportional to
\begin{equation}
A\sim \frac{e^2}{s}j_{\mu}J^{\mu},
\end{equation}
where $e$ is the charge of the electron, $s$ is the square of the
center-of-mass energy, $j_{\mu}$ is the $e^+e^-$ current, 
and $J^{\mu}$ is the hadronic current for the final
state. The object of many 
experiments is to measure the matrix elements of $J^{\mu}$. 
In $e^+e^-\rightarrow p\bar{p}$, a pair of spin-1/2
baryons with internal structure are produced, and the current
contains two independent form factors, electric and
magnetic, $G_E(q^2)$ and $G_M(q^2)$~\cite{perkins}.

Understanding nucluon structure is one of the central problems of
hadronic physics. In the time-like region, two processes, $\ep$ and
$\pe$, are used to measure the proton form factors $G_E(q^2)$ and
$G_M(q^2)$ as functions of the four-momentum transfer $q^2$. Data
samples in previous experiments are
limited~\cite{Bardin}-\cite{E835}. For $6<s<8$ GeV$^2$,
there is no data up till now.

In this letter, we use the data from the upgraded Beijing Spectrometer
(BESII) at the Beijing Electron-Positron Collider (BEPC) 
covering the center-of-mass
energy of 2.0-3.0 GeV in 1999~\cite{R99} and the data at 2.2, 2.6, and
3.07 GeV in 2004 to measure the cross section of
$e^+e^-\rightarrow p\bar{p}$, and also determine 
the proton form factor in this energy range.

\section{BES detector}

BES is a conventional solenoidal magnet detector that is
described in detail in Ref.~\cite{bes}; BESII is the upgraded version
of the detector~\cite{bes2}. A 12-layer vertex
chamber (VC) surrounding the beam pipe provides trigger and coordinate
information. A forty-layer main drift chamber (MDC), located
radially outside the VC, provides trajectory and energy loss
($dE/dx$) information for charged tracks over $85\%$ of the
total solid angle.  The momentum resolution is
$\sigma _p/p = 0.017 \sqrt{1+p^2}$ ($p$ in $\hbox{\rm GeV}/c$),
and the $dE/dx$ resolution for hadron tracks is $\sim 8\%$.
An array of 48 scintillation counters surrounding the MDC  measures
the time-of-flight (TOF) of charged tracks with a resolution of
$\sim 200$ ps for hadrons.  Radially outside the TOF system is a 12
radiation length, lead-gas barrel shower counter (BSC).  This
measures the energies
of electrons and photons over $\sim 80\%$ of the total solid
angle with an energy resolution of $\sigma_E/E = 22\%/\sqrt{E}$ ($E$
in GeV).  Outside of the solenoidal coil, which
provides a 0.4~Tesla magnetic field over the tracking volume,
is an iron flux return that is instrumented with
three double layers of  counters that
identify muons of momentum greater than 0.5~GeV/$c$.

\section{Event selection}

To select $\ep$, 
the following criteria
are used:

\begin{enumerate}
\item There must be two oppositely charged tracks in the MDC. Each track
      should have a good helix fit in the polar angle range
      $|\cos\theta|<0.8$ in the MDC, and the point of closest approach of
      the tracks to the beam axis should be within 2 cm in the radial
      direction and within 15 cm of the interaction point longitudinally. 

\item Tracks should be back-to-back. %
Fig.~\ref{Acol} shows the acollinearity ($Acol$) distributions for Monte
Carlo (MC) simulated events at $\sqrt {s}$ = 2.0, 2.2, 2.6, and 3.0
GeV. Because of energy loss, the $Acol$ distribution for low energy
charged particles is somewhat different from higher energy
ones. For 2.0 GeV data, we require that $Acol$ is less than
$10^\circ$, and for other energy points, less than $3^\circ$.

\item To remove Bhabha events, $E_p/p_p<0.6$ is required, where $E_p$
is the deposited energy in the BSC and $p_p$ is the momentum of the candidate
proton. Fig.~\ref{scengp} shows $E_p/p_p$ distributions for
protons and positrons for MC $\EE \rightarrow p\pbar$ events and
Bhabha events at $\sqrt {s}$ = 2.2, 2.6, and 3.0 GeV.
This requirement removes most Bhabha background.  

\begin{figure}[htbp]
\centerline{\psfig{figure=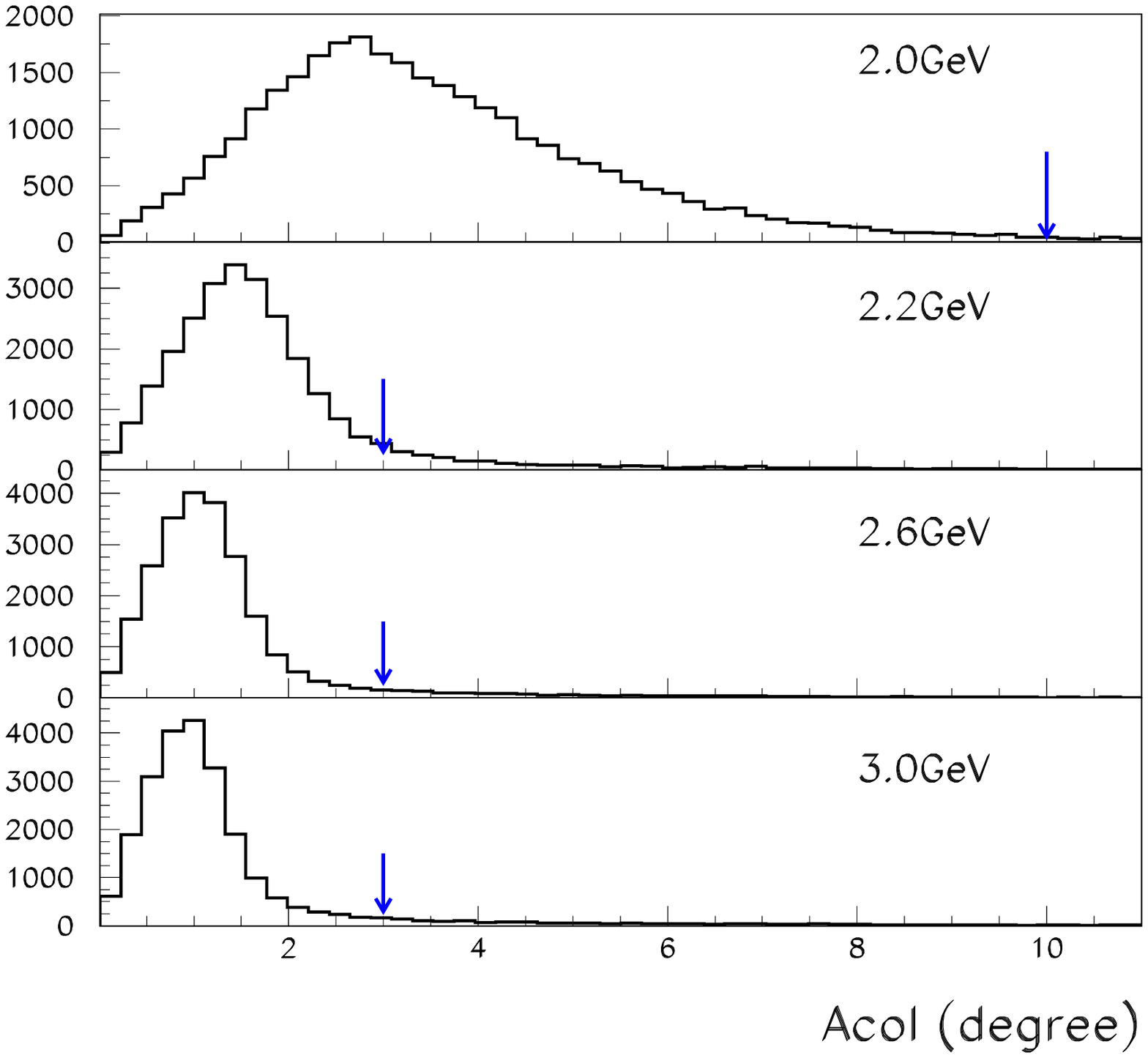,width=7cm,angle=0}}
\caption{$Acol$ distributions for MC $\ep$ at $\sqrt {s}$ = 2.0, 2.2, 2.6, and
3.0 GeV. The arrows show the selection requirement for each case.}
\label{Acol}
\end{figure}

\begin{figure}[htbp]
\centerline{\psfig{figure=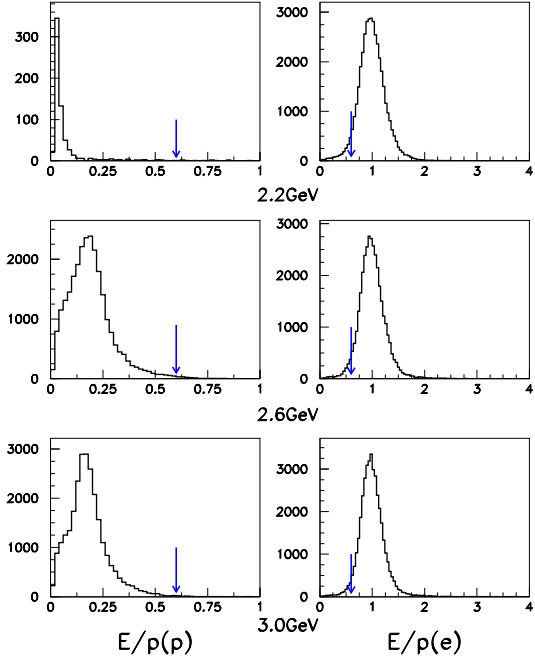,width=7cm,angle=0}}
\caption{Distributions of the ratio of the deposited energy in the BSC
  and the momentum for
final state particles of MC $\EE \rightarrow p\pbar$ events and
Bhabha events at $\sqrt {s}$  = 2.2, 2.6, and 3.0 GeV. Left plots are
protons, right are positrons.}
\label{scengp}
\end{figure}

\item The momentum is required to be within $3 \sigma_p$ of the
  nominal proton (antiproton) momentum for $\ep$ at each energy
  point, where $\sigma_p$ is the momentum resolution.

\item Lastly, $dE/dx$ information is used to identify $\ppb$ pairs at
$\sqrt {s}$ = 2.0 and 2.2 GeV, and TOF information is used at all
other energy points~\cite{lihh}.  Each charged track must satisfy
$Prob_{p}>0.01$, where $Prob_{p}$ is the particle identification
confidence level for the proton or antiproton hypothesis determined
either from $dE/dx$ or TOF information.

\end{enumerate}

The numbers of events passing the selection criteria are listed in Table ~\ref{result}.

\begin{table*}[htbp]
\begin{center}
\caption{Summary of results at center-of-mass energies
from 2.0 to 3.07 GeV.
N is the number of $\ep$ events,
$\cal L$ is the integrated luminosity,
$\varepsilon$ is the detection efficiency,
$F_\varepsilon$ is the detection efficiency correction factor from the
PID
efficiency difference between the MC sample and  the real data,
$1+\delta$ is the initial state radiation correction factor,
$F_c$ is the correction factor for the Coulomb effect,
$F_f$ is the final state radiation correction factor,
$\sigma_0$ is the measured lowest order cross section, and
$|G|$ is the form factor. In the last two columns,
the first error is
statistical and the second is systematic.
}
\label{result}
\vspace{0.5cm}
\begin{tabular}{cccccccccc}\hline
$\sqrt{s}$ &N &$\cal L$ &$\varepsilon$ & $F_\varepsilon$& $1+\delta$ &$F_c$
&$F_f$ &$\sigma_0$ &$|G|$ \\
(GeV)&   & $(nb^{-1})$&     &  & & & & (pb) & ($\times 10^{-3}$)  \\ \hline
2.0& $3^{+2.3}_{-1.9}$ &  $45.8\pm 1.4$& $0.53\pm 0.03$& 0.41&   0.89& 1.03&  0.99&
$330^{+253}_{-209} \pm 24$& $175 ^{+67}_{-55} \pm 6$ \\
2.2& $29 \pm 5.4$      & $123.5\pm 3.7$& $0.56\pm 0.02$ & 1.04&  0.98& 1.02&   0.99&
$408\pm 76  \pm 22$& $179 \pm 17 \pm 5$ \\
2.4& $2^{+2.2}_{-1.3}$ &  $61.0\pm 1.6$& $0.48\pm 0.02$& 1.03&  1.04& 1.02&   0.98&
$64 ^{+73}_{-41}   \pm 4$ & $72  ^{+41}_{-23}  \pm 2$ \\
2.5& $5^{+2.8}_{-2.2}$ &  $47.0\pm 1.0$& $0.50\pm 0.02$& 0.99&  1.07& 1.02&   0.98&
$201^{+113}_{-91}   \pm 13$& $131 ^{+37}_{-29} \pm 4$ \\
2.6& $24 \pm 4.9$      &  $1351\pm 24$ & $0.51\pm 0.02$& 0.96&  1.10& 1.02&   0.98&
$33 \pm 7   \pm 2$ & $54  \pm 6  \pm 2$ \\
2.7& $2^{+2.2}_{-1.3}$ &  $71.6\pm 2.1$& $0.48\pm 0.02$& 1.00&  1.13& 1.02&   0.98&
$51 ^{+58}_{-32}    \pm 5$ & $70  ^{+39}_{-22} \pm 3$ \\
2.8& $2^{+2.2}_{-1.3}$ &  $89.0\pm 1.8$& $0.50\pm 0.02$& 0.96&  1.17& 1.02&   0.98&
$40 ^{+45}_{-25}    \pm 4$ & $63  ^{+36}_{-20} \pm 3$ \\
2.9&  0&  $94.0\pm 2.6$& $0.49\pm 0.02$& 0.96&  1.20&   1.02& 0.98& $<$51&$<$73 \\
3.0& $4^{+2.8}_{-1.7}$ &  $947\pm 22$  & $0.50\pm 0.02$& 0.96&  1.24& 1.01&   0.98&
$7  ^{+5}_{-3}      \pm 1$ & $28  ^{+10}_{-6}  \pm 1$ \\
3.07&$9^{+3.8}_{-2.7}$ &  $2347\pm 59$ & $0.49\pm 0.02$& 0.96&  1.27& 1.01&   0.98&
$7  ^{+3}_{-2}      \pm 1$ & $27  ^{+6}_{-4}  \pm 1$ \\
\hline
\end{tabular}
\end{center}
\end{table*}

\section{Luminosity}

The integrated luminosity is determined from large-angle Bhabha
events using~\cite{lum1}\cite{lum2}
\begin{equation}
{\cal L}=\frac{N_{ee}}{\varepsilon^{ee}_{trg}\cdot A \cdot
C_{\varepsilon}\cdot \sigma_{ee}},
\end{equation}
where, $N_{ee}$ is the number of Bhabha events selected using BSC
information only, $\varepsilon^{ee}_{trg}$ is the trigger efficiency
for Bhabha events, $A$ is the acceptance of Bhabha events estimated by
MC simulation using the same selection criteria as for the data, $C_{\varepsilon}$
is the efficiency correction factor, which is used to correct for
differences between the MC and data angular distributions due to the
ribs in the BSC, and $\sigma_{ee}$ is the Bhabha cross section.

\section{Efficiency}

A MC simulation is used for the determination of the detection
efficiency. In the $e^+e^- \rightarrow p \overline p (\gamma)$
generator, corrections for initial state radiation~\cite{kureav}, the
Coulomb effect~\cite{Coulomb}\cite{tau}, and final state radiation
~\cite{final} have been taken into account. The correction
factors for these items, $1+\delta$, $F_c$, and $F_f$, respectively,
are listed in Table~\ref{result}.

For each energy point, 50,000 MC events are generated. MC events
must satisfy the same selection criteria as used for the real data. The
detection efficiencies, $\varepsilon$, are given in
Table~\ref{result}.

\section{Systematic errors}

Systematic errors come from uncertainties in the detection efficiency,
trigger efficiency, luminosity, and background contamination. The detection
efficiency uncertainty includes the MC
statistical error and the differences in the particle identification
(PID) and  tracking efficiencies for the MC sample and the real data.

There are few  $\ep$ events, so $p$ and $\bar{p}$ samples from
$J/\psi \rightarrow \pi^+\pi^-p\bar{p}$ are used for the PID
efficiency study.  The momenta of the candidate $p$ and $\bar{p}$
tracks are required to be within 30 MeV/c of the expected values for
each energy point. The samples are obtained without using PID.  The
PID efficiency for each energy point is then detemined by the fraction
of p and $\bar{p}$ tracks that pass PID selection criteria. The
same method is used for MC $J/\psi \rightarrow \pi^+\pi^-p\bar{p}$
events to determine the PID efficiency for the MC data.  At 2.0 GeV, there
is a large difference in PID efficiency between the MC sample and the data, so
for all energy points detection efficiencies ($F_\varepsilon$) are
corrected for this difference, and the errors in the PID efficiency
difference are taken as a source of systematic error.

For $\ep$, possible backgrounds are from  $\EE \rightarrow
\EE (\gamma)$, $\MM (\gamma)$, $\pp$, $\kk$, and $p\pbar\pi^0$. MC events are
generated
for these five
decays at $\sqrt{s}=$ 2.2, 2.6 and 3.0 GeV to estimate the
amount of background contamination, which
is
included as a systematic error: $1.5\%$ for 2.0 and 2.2
GeV, $4.4\%$ for 2.4, 2.5, and 2.6 GeV, and $7.8\%$ for
other energy points.

Systematic errors are listed in Table~\ref{sys}. An uncertainty of 1.0$\%$ is taken 
for the initial state radiation correction. 

\begin{table*}[htbp]
\begin{center}
\caption{The relative systematic error ($\%$):
$\Delta\varepsilon/\varepsilon$ is the contribution from detection
efficiency, including the  MC statistical error ($err_N$), the PID difference
($err_{PID}$) and tracking efficiency difference ($err_{track}$) between MC 
and real data; $\Delta\varepsilon_{trig}/\varepsilon_{trig}$ is the
contribution from the trigger efficiency; $\Delta{\cal L}/{\cal L}$ is the
contribution from luminosity; and BG is the contribution from the background
contamination.}
\label{sys}
\vspace{0.5cm}
\begin{tabular}{ccccccccc}\hline 
$\sqrt{s}$ &\multicolumn{4}{c}{$\Delta\varepsilon/\varepsilon$}&
$\Delta\varepsilon_{trig}/\varepsilon_{trig}$ &$\Delta{\cal L}/{\cal L}$&
BG &total \\ \cline{2-5}
(GeV) &$err_N$ &$err_{PID}$ &$err_{track}$ &total & & & & \\ \hline
2.0 &0.4 &5.0 &4.0 &6.4 &0.5 &3.0 &1.5  & 7.3 \\
2.2 &0.4 &0.6 &4.0 &4.1 &0.5 &3.0 &1.5  & 5.3 \\
2.4 &0.5 &0.9 &4.0 &4.1 &0.5 &2.7 &4.4  & 6.6 \\
2.5 &0.4 &1.0 &4.0 &4.1 &0.5 &2.2 &4.4  & 6.4 \\
2.6 &0.4 &1.0 &4.0 &4.1 &0.5 &1.8 &4.4  & 6.3 \\
2.7 &0.5 &1.0 &4.0 &4.2 &0.5 &2.9 &7.8  & 9.3 \\
2.8 &0.4 &1.9 &4.0 &4.4 &0.5 &2.0 &7.8  & 9.2 \\
2.9 &0.5 &1.9 &4.0 &4.4 &0.5 &2.8 &7.8  & 9.4 \\
3.0 &0.4 &1.9 &4.0 &4.4 &0.5 &2.3 &7.8  & 9.3 \\ 
3.07&0.5 &1.9 &4.0 &4.4 &0.5 &2.5 &7.8  & 9.3 \\ \hline
\end{tabular}
\end{center}
\end{table*}

\section{Results and summary}

The total cross section is determined from
\begin{equation}
\sigma_T=\frac{N}{{\cal L}\cdot \varepsilon \cdot \varepsilon_{trig}},
\label{sigmat}
\end{equation}
where N is the number of $\ep$ events, $\cal L$ is the integrated luminosity, 
$\varepsilon$ is the detection efficiency, and $\varepsilon_{trig}$ is the
trigger efficiency.
The lowest order cross section is determined from
\begin{equation}
\sigma_0=\frac{\sigma_T}{(1+\delta) \cdot F_c \cdot F_f},
\label{sigma0}
\end{equation}
where $1+\delta$, $F_c $, and $F_f$ are correction factors for
initial state radiation, the Coulomb effect, and final state radiation, respectively. 

The form factor can be calculated from the theoretical lowest order cross
section~\cite{Antonelli}
\begin{equation}
\sigma_{0}=\frac{4\pi\alpha^2\nu}{3s}(1+\frac{2m^2_p}{s})|G(s)|^2,
\label{sigma0_theory}
\end{equation}
in which $\alpha$ is the fine structure constant, 
$\nu$ is the proton velocity, $m_p$ is the proton mass, and $|G|$
is the form factor assuming $|G_E|=|G_M|$.

The trigger
efficiency $\varepsilon_{trig}$ for the hadron is about 100$\%$, and the
error is estimated to be 0.5$\%$~\cite{trig}. 

The cross section of $\ep$ and proton form factor have been measured
for 10 center-of-mass energies between 2.0 and 3.07 GeV.
The measured values are listed in Table~\ref{result}.
There is no signal found at $\sqrt{s}=2.9$ GeV, but upper limits on the
cross section and the form factor at the $90\%$ C.L. are given, using the method
from Ref.~\cite{stat1}.

\begin{figure}[htbp]
\centerline{\psfig{figure=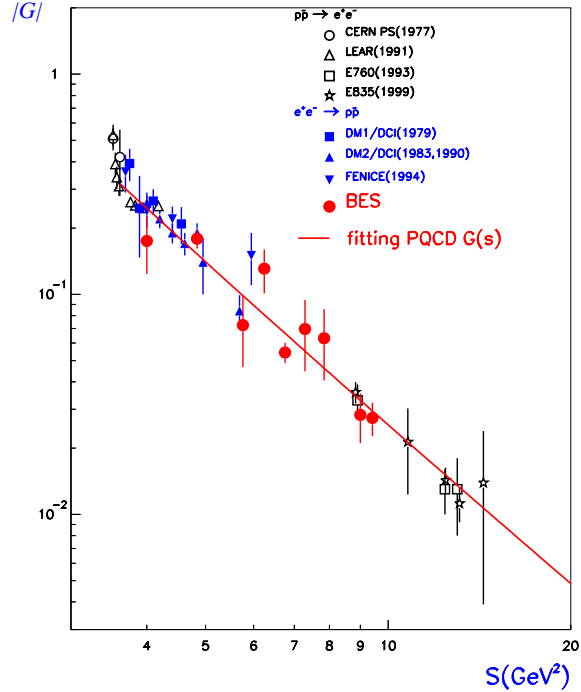,width=7.5cm,angle=0}}
\caption{Form factors measured by BES and other experiments. The line shows the 
energy dependence of $|G(s)|$ by fitting
all measurements.}
\label{gem} 
\end{figure}

For large momentum transfers
pQCD predicts~\cite{pqcd} that 
$q^4G$
should be nearly proportional to the square of the running coupling constant
for strong interactions, $\alpha_s^2(q^2)$, yielding the relation
\begin{equation}
|G|=\frac{C}{s^2ln^2(s/\Lambda^2)},
\label{gemfun}
\end{equation}
where $\Lambda=0.3$ GeV is the QCD scale parameter and C is a free parameter.
In Fig.~\ref{gem}, BES results are compared with other experimental
proton form factor results. The line is the $|G(s)|$ energy dependence
obtained by fitting all measurements with Eq.~\ref{gemfun},
and the result is consistent with the pQCD prediction.

\section*{Acknowledgments}

     The BES collaboration thanks the staff of BEPC for their hard
efforts. This work is supported in part by the National Natural
Science Foundation of China under contracts Nos. 10491300,
10225524, 10225525, 10425523, the Chinese Academy of Sciences under
contract No. KJ 95T-03, the 100 Talents Program of CAS under
Contract Nos. U-11, U-24, U-25, and the Knowledge Innovation
Project of CAS under Contract Nos. U-602, U-34 (IHEP), the
National Natural Science Foundation of China under Contract No.
10225522 (Tsinghua University), and the Department of Energy under
Contract No.DE-FG02-04ER41291 (U Hawaii).

\end{document}